\documentclass[a4paper,12pt]{article}

\usepackage[english]{babel}
\usepackage[utf8]{inputenc}
\usepackage{amsmath}
\usepackage{graphicx}
\usepackage[colorinlistoftodos]{todonotes}
\usepackage[margin=1in]{geometry}
\usepackage[hidelinks]{hyperref}
\usepackage{amssymb}
\usepackage{stix}
\usepackage{gensymb}
\usepackage {tikz}
\usepackage{adjustbox}
\usepackage{multirow}
\usepackage{graphicx}
\usepackage{xcolor}
\usepackage{caption}
\usepackage{subcaption}
\usepackage{graphicx}
\usepackage{algorithm}
\usepackage{algpseudocode}
\usepackage{tabularx}

\usepackage{array}

\usetikzlibrary {positioning}

\title{Scenario Discovery for Urban Planning: The Case of Green Urbanism and the Impact on Stress}
\bigskip
\usepackage{float}
 \author{\textit{Lorena Torres Lahoz$^{1*}$}, \textit{Carlos Lima Azevedo$^{1}$},
 \textit{Leonardo Ancora $^{2}$},\\
 \textit{Paulo Morgado $^{2}$}, 
\textit{Zenia Kotval $^{3}$},
 \textit{Bruno Miranda $^{4}$},
 \textit{Francisco Camara Pereira $^{1}$}, 
 \\\\
\textit{$^1$DTU Management, Technical University of Denmark}\\
\textit{$^2$Associate Laboratory TERRA, Centre of Geographical Studies,}\\
\textit{Institute of Geography and Spatial Planning, University of Lisbon, Lisbon, Portugal
}\\
\textit{$^3$ School of Planning, Design and Construction, Michigan State University
}\\
\textit{$^4$ Faculty of Medicine, University of Lisbon
}\\
\textit{$^*$Corresponding author: ltola@dtu.dk}
}
\begin{document}
\maketitle
\section*{Abstract}
% This paper explores Scenario Discovery (SD), a framework for decision-making under uncertainty, in urban planning. We demonstrate its application through a case study to reduce stress levels in outdoor walking by exploring the relationship between urban environments and subjective well-being. We analyse how urban features such as greenery influence stress levels using a predictive model based on emotional responses collected from a neuroscience-based outdoor experiment in Lisbon. Combining these insights with detailed urban data from Copenhagen, we then explore the use of Scenario Discovery to identify conditions under which specific urban feature interventions, such as increasing vegetation, are most effective under different sources of uncertainty. The analysis revealed that different urban areas require tailored strategies to reduce stress, particularly in scenarios involving increased urban density and changing population characteristics. This work showcases Scenario Discovery as a systematic approach for identifying robust policy pathways in urban planning, opening the door for its exploration in other urban decision-making contexts where uncertainty and design resiliency are critical.

Urban environments significantly influence mental health outcomes, yet the role of an effective framework for decision-making under deep uncertainty (DMDU) for optimizing urban policies for stress reduction remains underexplored. While existing research has demonstrated the effects of urban design on mental health, there is a lack of systematic scenario-based analysis to guide urban planning decisions. This study addresses this gap by applying Scenario Discovery (SD) in urban planning to evaluate the effectiveness of urban vegetation interventions in stress reduction across different urban environments using a predictive model based on emotional responses collected from a neuroscience-based outdoor experiment in Lisbon. Combining these insights with detailed urban data from Copenhagen,  we identify key intervention thresholds where vegetation-based solutions succeed or fail in mitigating stress responses. Our findings reveal that while increased vegetation generally correlates with lower stress levels, high-density urban environments, crowding, and individual psychological traits (e.g., extraversion) can reduce its effectiveness. This work showcases our Scenario Discovery framework as a systematic approach for identifying robust policy pathways in urban planning, opening the door for its exploration in other urban decision-making contexts where uncertainty and design resiliency are critical.

\newpage
\section{Introduction}
Uncertainty can be defined as 'limited knowledge about future, past or current events, and it can be quantified as any departure from the ideal of complete determinism' \cite{Uncertainty}.
It is often quantified in terms of “probability”. For example, in the context of decision-making, uncertainty is measured by identifying plausible future conditions or scenarios, where probabilities are assigned to different outcomes based on their likelihood. Another example is when the input data is incomplete or noisy, leading to uncertainty in model predictions, where probabilistic measures are used to estimate the confidence or likelihood of various predicted outcomes. 

In this paper, we will focus on deep uncertainty. Lempert et al. \cite{sd}  have defined it as “the condition in which analysts do not know or the parties to a decision cannot agree upon (1) the appropriate models to describe interactions among a system’s variables, (2) the probability distributions to represent uncertainty about key parameters in the models, and/or (3) how to value the desirability of alternative outcomes.”
Deep uncertainty refers to situations where there is substantial ambiguity and disagreement regarding the appropriate models to use, the probabilities of different outcomes, and even the criteria for evaluating those outcomes. This phenomenon often arises in complex systems where the information is incomplete or overly intricate, making it challenging to make predictions.

Robust Decision-Making (RDM) is a quantitative decision-analytic method that utilizes available information to assist decision-makers in identifying more effective strategies for achieving their goals in the face of deep uncertainties. The primary objective of RDM is to discover policies that perform well in a wide range of potential future scenarios, rather than seeking a single optimal policy for a specific future scenario, which is common in traditional optimization frameworks \cite{SD1}. 
A crucial step in the Robust Decision-Making process is identifying the combinations of input parameters in a simulation model for which a proposed robust policy underperforms compared to alternative strategies. This procedure is known as Scenario Discovery (SD).

Scenario Discovery helps policymakers and analysts identify policy-relevant scenarios. It defines scenarios as sets of possible future states of the world that highlight \emph{vulnerabilities} in proposed policies. These scenarios are described in ways that enable decision-makers and stakeholders to easily understand and use them.
In this context, the concept of vulnerability refers to future states in which a proposed policy may fail to achieve its performance goals, that is, when its outcomes deviate significantly from those of an optimal policy.
Scenario Discovery employs a participatory computer-assisted process to support Robust Decision-Making (RDM). Rather than predicting the future, scenarios aim to "bound it" \cite{sd}. The ultimate goal is to help stakeholders with diverse perspectives identify and prioritize the combinations of key drivers most critical to future planning. Unlike trend-based forecasting, discovery scenarios focus on understanding uncertainties, identifying emergent patterns, and testing the resilience of urban policies against a range of possible developments \cite{Zenia1} \cite{Zenia2}.
It also facilitates the evaluation of trade-offs between economic growth and environmental conservation, helping planners integrate sustainability principles into long-term urban development strategies \cite{Zenia3}.

Traditionally, Scenario Discovery is performed using sparse sampling density methods, which could easily limit the statistical significance of the resulting scenarios and are pretty computationally expensive. Some studies have pointed out that a clear improvement would be to employ adaptive sampling methods to generate more simulation models near the scenarios' boundaries to improve confidence in the statistical significance of the results \cite{sd}.
Thus, we explore new machine learning approaches based on active learning \cite{AL} that address sampling density challenges by iteratively requesting new simulation model runs to add cases to the dataset in those most needed regions to improve the performance and efficiency of the scenario discovery algorithms.

In this paper, we present a case study applied in the city of Copenhagen to reduce stress under a wide range of future uncertainties; our aim is to propose a practical and replicable application of Scenario Discovery that can be easily used and understood by policymakers to make Scenario Discovery usage more accessible and practical in urban planning while highlighting their benefits in the planning process.
Urban planners employ discovery scenarios to investigate alternative trajectories of urban growth, technological change, and socio-economic transitions. The process often involves constructing multiple scenarios based on varying assumptions about demographic shifts, economic transformations, governance models, and environmental conditions \cite{Zenia4}. These scenarios allow planners to critically assess the long-term implications of different urban policies and explore how cities can adapt to dynamic challenges \cite{Zenia5}.
Moreover, one of the main strengths lies in their ability to foster participatory planning. By involving diverse stakeholders, researchers and planners it can integrate multiple perspectives into decision-making processes \cite{Zenia6}.

This paper is structured as follows: the next section presents a review of the literature on both RDM and the algorithms employed to perform the scenario discovery process. Section 3 covers the methodologies proposed. Section 4 presents the case study in which these methodologies are applied, and Section 5 analyzes the results. Finally, the paper finishes with Section 6, which includes the conclusions and limitations of the present paper.

\section{Literature Review}
\subsection{Handeling uncertainty in Urban Planning}
Traditionally, there have been four (not mutually exclusive) ways of dealing with deep uncertainty in policymaking \cite{deep uncertainty}. \textit{Resistance} implies planning for the worst-case scenario, an approach that can be very costly and yet unsuccessful in handling Black Swans \cite{Black Swan}. \textit{Resilience}, apply policies that can recover quickly given any disturbance. It accepts a small quantity of loss initially but focuses on a quick recovery. \textit{Static robustness}, implement policies that will perform reasonably well in most of the possible future cases. A robust policy is good enough across various future case scenarios, as opposed to an optimal policy with the best performance for a one-case scenario. \textit{Adaptative robustness}, which combines a robust policy with a monitoring system. When some of the monitored values reach a defined threshold, the policy can be changed during the implementation phase to be more suitable, given the changes in the initial conditions. While this may be the most attractive option, it is often infeasible to design.
This paper focuses on static robustness within the RDM framework employing Scenario Discovery techniques.

The inherent complexity of urban environments and their uncertain future conditions necessitate exploring innovative approaches and tools to assist the current planning practices.
The problem of dealing with uncertainty is further amplified by the fact that cities are complex (self) adaptative systems \cite{ref1}, \cite{ref2} and urban planners often lack awareness of uncertainty or the knowledge to manage effectively. The rapid changes of the past decade have further exposed the limitations of long-term predictions in a time of increasing uncertainty and thus the need for new ways to approach long-term planning. A workflow utilized by urban planners to address the uncertainty about cities’ future and inherent complexity using Exploratory Analysis and Scenario Discovery was presented in a case study in Rome, Italy \cite{Urban2}.
While computational scenario planning holds promise as a tool for addressing uncertainties, its successful implementation often requires interdisciplinary collaboration \cite{BookUrban}. An example of this interdisciplinary collaboration can be seen in an integrated framework that combines Cellular Automata (CA) models with exploratory modelling to assess how climate uncertainties affect long-term land-use changes \cite{Urban1}. In another study, a simple cohort model was used to explore uncertainties, where they discovered that the success of the plan depends more on economic and energy factors than on the transport policy itself, demonstrating that even a limited application of RDM can usefully illuminate the vulnerabilities of a plan \cite{Urban3}.
However, further research is needed to generalize the integration of resilience into urban planning under uncertainty \cite{Urban5}. The application of DMDU approaches in the built environment is still limited and requires further studies. 

On the other hand, some studies have focused on the relationship between the urban environment and psychological stress, a term called Neurorubanism \cite{Paulo1}, which combines neuroscience and urban planning to promote healthier cities through evidence-based policies. Relying on experiments using wearable physiological sensors and self-reported data, they identify spatial patterns in urban areas that correspond to stress responses, proving an existing relationship between the urban environment and our emotions \cite{Paulo2}, \cite{Paulo3}.

Therefore, this study showcases the first example of a RDM framework as a systematic approach for identifying robust policy pathways in urban planning applied to a neurorubanism case, opening the door for its exploration in other urban decision-making contexts where uncertainty and design resiliency are requiered.

\subsection{Scenario Discovery}
Scenario Discovery (SD) aims to provide summarized, accessible, and actionable insights for decision-makers on the key vulnerabilities of a given policy across possible future states of the system. In addition, these principles can be extended to explore other relevant policies. For instance, Scenario Discovery can help identify policies that perform exceptionally well compared to other alternatives, or focus on future scenarios that are more probable based on a known exogenous distribution. In essence, Scenario Discovery addresses the question: \textit{``What are the most critical vulnerabilities of the strategy under consideration?''} Vulnerable cases are defined as combinations of uncertain input parameters that result in policy outcomes failing to meet performance requirements.

Therefore, the primary goal of Scenario Discovery is to identify combinations of the input parameters of the simulation model that are strongly correlated with vulnerabilities. A simulation model is a simplified version of a real-world system that helps us understand how it works or predict what might happen in different future situations. These combinations, referred to as descriptors, define a subspace within the uncertainty input space, commonly known as a "box" or a "scenario".

In order to accomplish its objective, Scenario Discovery results are based on three metrics:
\begin{itemize}
  \item \textbf{Coverage}: the percentage of vulnerable points captured in a box with respect to the total number of vulnerable cases in the whole exploration space.
  \item \textbf{Density}: the fraction of vulnerable cases within the box relative to the total number of points inside the box.
  \item \textbf{Interpretability}: indicates how easy it is to understand a scenario, usually described by the number of constraints that define each box.
\end{itemize}  
An ideal set of scenarios would combine high density, coverage and interpretability \cite{Algorithms}. However, these measures are generally interrelated since an increase in coverage will usually increase the number of captured vulnerable cases but decrease the box's density.
\begin{figure}[H]
    \centering
    \includegraphics[width = 0.5
    \textwidth]{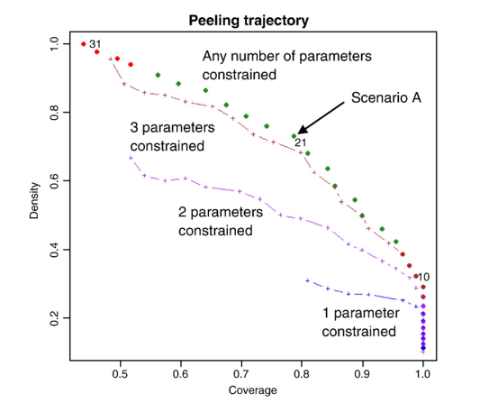}
    \caption{Example of coverage/density trade-off curves using 1,2,3, or any number of restricted parameters \cite{SD}}
    \label{QD}
\end{figure}
In a nutshell, Scenario Discovery consists of a trade-off between optimization and exploration that highlights uncertain input conditions where the proposed policy will not satisfy a particular success/failure criterion, a crucial step in the design of robust policies. For example, we could introduce a policy to decrease the price of public transport to increase travel accessibility. Although this policy may seem promising, its effectiveness could diminish if the general income of the population also decreases.

\subsubsection{Algoritims for Scenario Discovery}
To our knowledge, no algorithms have been designed specifically to perform the tasks required for
scenario discovery \cite{Algorithms}. However, the Patient Rule Induction Method (PRIM) and the Classification and Regression Tree (CART) are the most traditionally employed.

PRIM \cite{PRIM} is designed to identify regions in the input space where the mean output value is greater than the overall mean output across the entire space. To achieve this objective, PRIM employs a process known as "peeling." The algorithm begins with a box that covers the entire input space. At each step, a small sub-box is removed from the edges of the current box. The sub-box selected for removal is the one that results in the highest mean output value within the remaining box. This process is repeated until the support of the final box, defined as the number of vulnerable points divided by the total number of points, reaches a specified threshold. 

\begin{figure}[H]
    \centering
    \includegraphics[width = 0.8
    \textwidth]{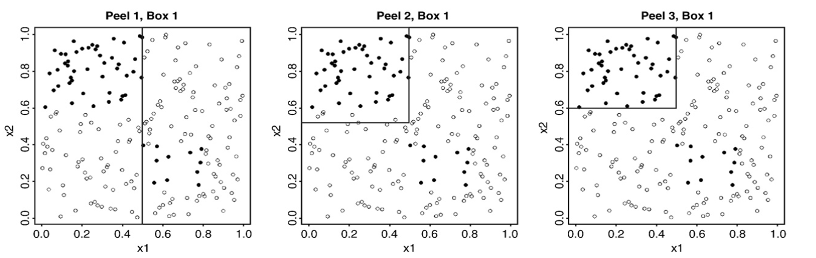}
    \caption{An example of a sequence of the peeling processes performed by the PRIM algorithm on a given dataset \cite{SD}}
    \label{PRIM}
\end{figure}

Figure \ref{PRIM} illustrates a series of peeling operations applied to a given dataset, where the bold points indicate vulnerable cases (represented as 1), while the non-vulnerable cases are marked as 0.

PRIM (Patient Rule Induction Method) includes a hyperparameter called patient, which controls the percentage of data points removed during each iteration of the peeling process. While smaller values of the patient are generally preferred, they can sometimes lead to the algorithm prematurely truncating the ends of the boxes, which might otherwise span the entire parameter range.

The peeling process can be repeated multiple times. After constructing the first box, a second box can be generated by using the initial data points and excluding the observations already covered by the first box. This iterative process, known as covering, facilitates a more thorough exploration of the entire input space.

On the other hand, CART (Classification and Regression Trees) aims to minimize misclassification errors, producing regions of the input space with high purity. The algorithm outputs a decision tree that classifies cases (e.g., vulnerable or not vulnerable) based on combinations of input dimensions \cite{Algorithms}. CART includes a hyperparameter called pruning, which simplifies the decision tree for better interpretability and predictive performance. Initially, the algorithm creates a large, complex tree by dividing the input space extensively. Pruning then combines the final splits based on a selected criterion (such as the minimum number of points in a split), archiving a simpler and more interpretable tree.

In summary, CART divides the entire input space into disjoint regions, unlike PRIM, whose boxes can overlap during the covering process. While CART often achieves better coverage and density performance than PRIM, studies have generally found PRIM more useful for policy analysis. This is because CART frequently requires an infeasibly large number of regions to achieve sufficient coverage, compromising interpretability \cite{Algorithms}.

Finally, user interaction can enhance the performance of both algorithms. In CART, users can select from different pruned trees, while in PRIM, they can choose the most suitable combination of coverage, density, and interpretability based on specific criteria or domain knowledge.

Although efforts have been made to enhance traditional algorithms—such as applying PRIM after orthogonal rotations in the dataset \cite{OrtogonalSD}, a promising improvement lies in the use of adaptive sampling methods. These methods generate additional simulation model runs near the edges of scenarios, increasing confidence in the statistical significance of the results.
Emerging machine learning approaches, particularly those based on active learning \cite{AL-SD}, can address sampling density challenges more effectively. By iteratively selecting new simulation model runs in regions where additional data is most needed, these techniques can enhance the performance of scenario discovery algorithms, resulting in more accurate and robust insights.

\section{Methodology}
In this section, we explain the Scenario Discovery framework employed in this case study, including our proposed modifications and the new Scenario Discovery algorithms for which we will give a proof-of-concept in the results.
\subsection{Scenario Discovery Framework}
The first step to apply the Scenario Discovery Framework is to define the problem based on the RAND’s XLRM framework, which determines the limits and scope of Scenario Discovery algorithm. In the XLRM framework \cite{XLRM}, X stands for Exogenous uncertainties (X), which are all the variables we have no control over; however, they play an essential part in the possible outcomes of our actions. Policy levers (L) are the actions policymakers can apply to modify the current environment. The Relationships (R) are the potential ways the future could evolve based on the policymakers’ choices of levers and the manifestation of the uncertainties. Given some initial conditions, this relationships could be seen as a simulation model. Finally, Measures (M) are the performance standards that policymakers and other interested stakeholders use to rank the desirability of various scenarios.

The second step is to sample points within the input uncertainty space. The most common approach is to use the Latin Hypercube Sampling (LHS) \cite{LHS} technique since it has one-dimensional uniformity, where for each input variable, its range is divided into the same number of equally spaced intervals as the number of observations, and there is exactly one observation in each interval. This technique ensures that the whole input space is initially explored uniformly.
The Latin Hypercube (LHS) provides a convenient experimental design for scenario discovery because it provides an efficient sample of a model's behaviour over the whole input space. However, LHS can be quite difficult and time-demanding for complex models.

The third step is to run the simulation model or the metamodel simulator in the selected input points, given a selected policy lever, and calculate their performance measure for each point. A metamodel is just a simpler version of a model that approximates a more complex and time-consuming model. A metamodel formulation is employed to predict the value of the distribution of vulnerable cases in a wide variety of regions without sampling all of them. One of the most common metamodel formulations is Gaussian Processes (GP) \cite{BO3}.

Finally, the fourth step consists in using the simulated observations to find the regions in the input space of uncertainties that highlight the most vulnerable cases for the considered policy levers.

\begin{figure}[H]
    \centering
    \includegraphics[width = 0.8
    \textwidth]{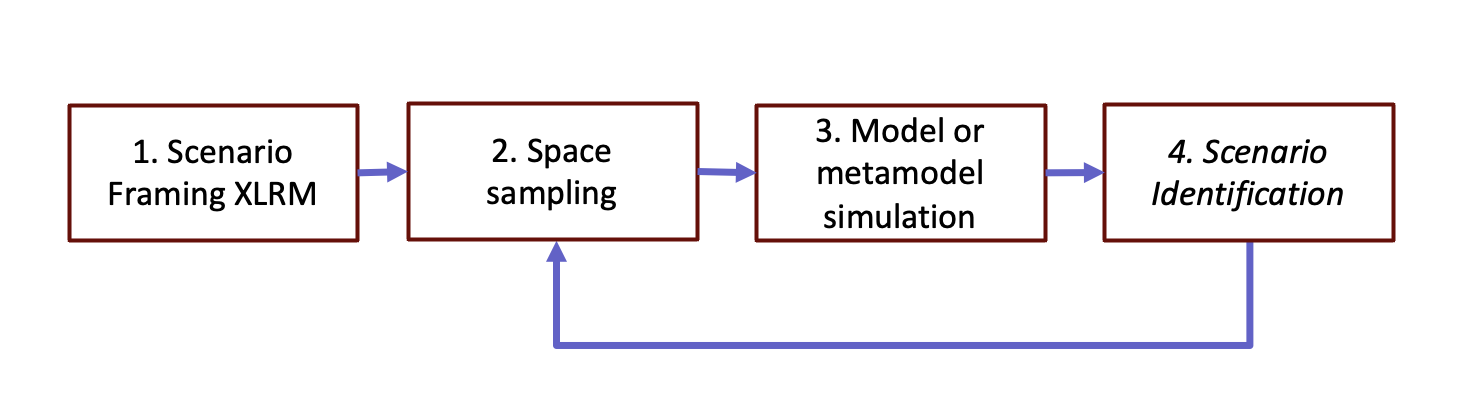}
    \caption{Proposed Scenario Discovery framework}
    \label{framework}
\end{figure}

In this paper, we consider a feedback loop from Step 4 back to Step 2 as shown in Figure \ref{framework}. Based on the scenario discovery results, we return to the sampling strategy to perform an adaptive sampling method that selects the next point based on a criterion related to the already found PRIM boxes.
This modification enables a more effective sampling that optimizes the knowledge of the already requested sampling points.
The following subsection presents different algorithms for performing this adaptative sampling strategy.
It is also worth noticing that the PRIM algorithm will be used for the Scenario Discovery proposed in all the presented strategies since PRIM has shown to be the most effective way of finding vulnerable and informative scenarios easily and efficiently.
\subsection{Perfomed algorithms}
\subsubsection {Baseline}
Firstly, some LHS samples need to be computed in the second step of our proposed framework. Therefore, to assess the optimal number of points needed with an LHS experimental design for Scenario Discovery proposes, a relative density metric, $J$, is proposed in \cite{density}. J is defined as,
\begin{equation}
J = {n_s}^{1/k}
\end{equation}
where $n_s$ is the number of runs to be carried out, and $K$ is the length of the input vector.

The optimal value of $J$, judged subjectively relative to other analyses in \cite{density}, is proposed to be at least between 1.5 and 3 or higher.

\subsubsection {Adaptative PRIM}
For the adaptative PRIM algorithm, we apply a combination of PRIM with targeted sampling, where a new box is constructed at each iteration of the algorithm, and a new point is sampled from the borders or the inside of the corresponding PRIM box at each step as shown in Algorithm \ref{Alg1}.
This formulation could greatly benefit Scenario Discovery in the urban context where simulation models are generally complex and time-consuming, and reducing simulation runs implies a great benefit.\\
For making adaptative sampling more time-efficiency, we employ a metamodel, specifically a Gaussian Process (GP), that belongs to the family of non-parametric kernel methods. This means that they do not assume any particular function form (as opposed to models linear regression for example, which expects a specific $y=w^Tx$ form) and that any prediction (e.g., for an input vector x) is entirely determined by the neighbourhood relationships (e.g. all vectors in the dataset closer to x  will contribute more to the prediction).
% Mathematically, a Gaussian Process is a collection of random variables, any finite number of which have joint Gaussian distributions. In other words, it is a multivariate Gaussian distribution with the same dimensions as the number of datapoints (vectors) in the space.
In addition, the posterior distribution is the function that results from incorporating the information on the previous points in our GP, and it is used for prediction. The posterior distribution of a Gaussian distribution is also Gaussian, which makes this estimation easier to compute and use.
\begin{algorithm} [H]
\caption{Adaptative PRIM}\label{Alg1}
\begin{algorithmic}
\State 1. Load an initial LHS input sample with its corresponding simulation output values $\mathcal{D} =$ \{${x_{init}, y_{init}}\}$ and fit a $\mathcal{GP}$ with the dataset $\mathcal{D}$. Load an LHS sample of input values $\mathcal{B}=\{x_{pos}\}$ where $|\mathcal{B}_x| > >|\mathcal{D}_x|$.
\For{$n_{iter}$}
\State a. Calculate the posterior $\mathcal{GP}$ in $\mathcal{B}$, $y_{pos}$,  and add it to the dataset $\mathcal{B} =\{x_{pos}, y_{pos}\}$.
\State b. Perform PRIM in $\mathcal{B}$.
\State c. Uniformly sample from the simulation model one or more points ($x_{new}$) from inside or on the borders of the selected PRIM box.
\State d. Compute the output of the selected points given the simulation model, $ \mathcal{D'} = \{x_{new}, {y_{new}}\}$, and add it to the previous dataset of simulated points $\mathcal{D} =  \mathcal{D} \cup \mathcal{D'}$.
\State 3. Fit a $\mathcal{GP}$ with the final dataset of simulation points $\mathcal{D}$.
\EndFor
\State 4. Calculate the posterior $\mathcal{GP}$ in $\mathcal{B}$ and perform PRIM to obtain the final boxes.
\end{algorithmic}
\end{algorithm}
where $n_{iter}$ is the number of times we pick new simulation points.
\subsubsection {Adaptative PRIM borders}
Similar to Adaptative PRIM, in Adaptative PRIM borders a new point is sampled uniquely from the borders of the corresponding PRIM box at each step, as shown in Algorithm 2, since we hypothesise that these are the points that carry more information about the location of the vulnerable points at each step.
\begin{algorithm} [H]
\caption{Adaptative PRIM borders}\label{Alg2}
\begin{algorithmic}
\State 1. Load an initial LHS input sample with its corresponding simulation output values $\mathcal{D} =$ \{${x_{init}, y_{init}}\}$ and fit a $\mathcal{GP}$ with the dataset $\mathcal{D}$. Load an LHS sample of input values $\mathcal{B}=\{x_{pos}\}$ where $|\mathcal{B}_x| > >|\mathcal{D}_x|$.
\For{$n_{iter}$}
\State a. Calculate the posterior $\mathcal{GP}$ in $\mathcal{B}$, $y_{pos}$, and add it to the dataset $\mathcal{B} =\{x_{pos}, y_{pos}\}$.
\State b. Perform PRIM in $\mathcal{B}$.
\State c. Uniformly sample from the simulation model one or more points ($x_{new}$) from the borders of the selected PRIM box.
\State d. Compute the output of the selected points given the simulation model, $ \mathcal{D'} = \{x_{new}, {y_{new}}\}$, and add it to the previous dataset of simulated points $\mathcal{D} =  \mathcal{D} \cup \mathcal{D'}$.
\State 3. Fit a $\mathcal{GP}$ with the final dataset of simulation points $\mathcal{D}$.
\EndFor
\State 4. Calculate the posterior $\mathcal{GP}$ in $\mathcal{B}$ and perform PRIM to obtain the final boxes.
\end{algorithmic}
\end{algorithm}
where $n_{iter}$ is the number of times we pick new simulation points.\\
These algorithms, together with the proposed Scenario Discovery framework, are applied and exemplified in our proposed case study in the next section.
\section{Case study: Reducing stress in Greater Copenhagen}
Stress is broadly defined as the body's nonspecific and predictable response to demands placed upon it, however stress becomes harmful when environmental demands exceed a person’s ability to cope \cite{Stress}. Chronic mental stress is associated with various physical and psychological conditions, including cardiovascular diseases, hypertension, diabetes, cancer, headaches, depression, anxiety, and insomnia \cite{Stress1}. The variability in stress responses is influenced by numerous factors, such as variations in individual characteristics and environmental contexts \cite{Stress2}. For example, exposure to light or air pollution or excessive noise has been linked to poorer mental health outcomes. Conversely, walking in urban vegetation has shown to have benefits to psychophysiological wellbeing \cite{Stress3}. Moreover, mixed-land use that fosters community interaction (e.g., shopping, recreation, volunteering), and well-connected, pedestrian-friendly environments have all been associated with improved mental and emotional well-being. \cite{21}

In this study, we showcase the application of Scenario Discovery in the exploration of urban feature intervention for reducing stress during urban walks.  We rely on a data set collected as part of the eMOTIONAL Cities project \cite{11}. In the lab experiment, participants were asked to watch first-person street walk videos that were carefully selected to capture different city environments. The environments were previously selected through participatory experiment design \cite{11} and recorded accordingly. Self-reported data on their arousal and valence levels was asked while watching the videos. The data was collected for 20 participants using videos from street environments from Lisbon, Portugal. More information on the experiment can be accessed here \cite{11}.

\begin{figure}[H]
    \centering
    \includegraphics[width = 1
    \textwidth]{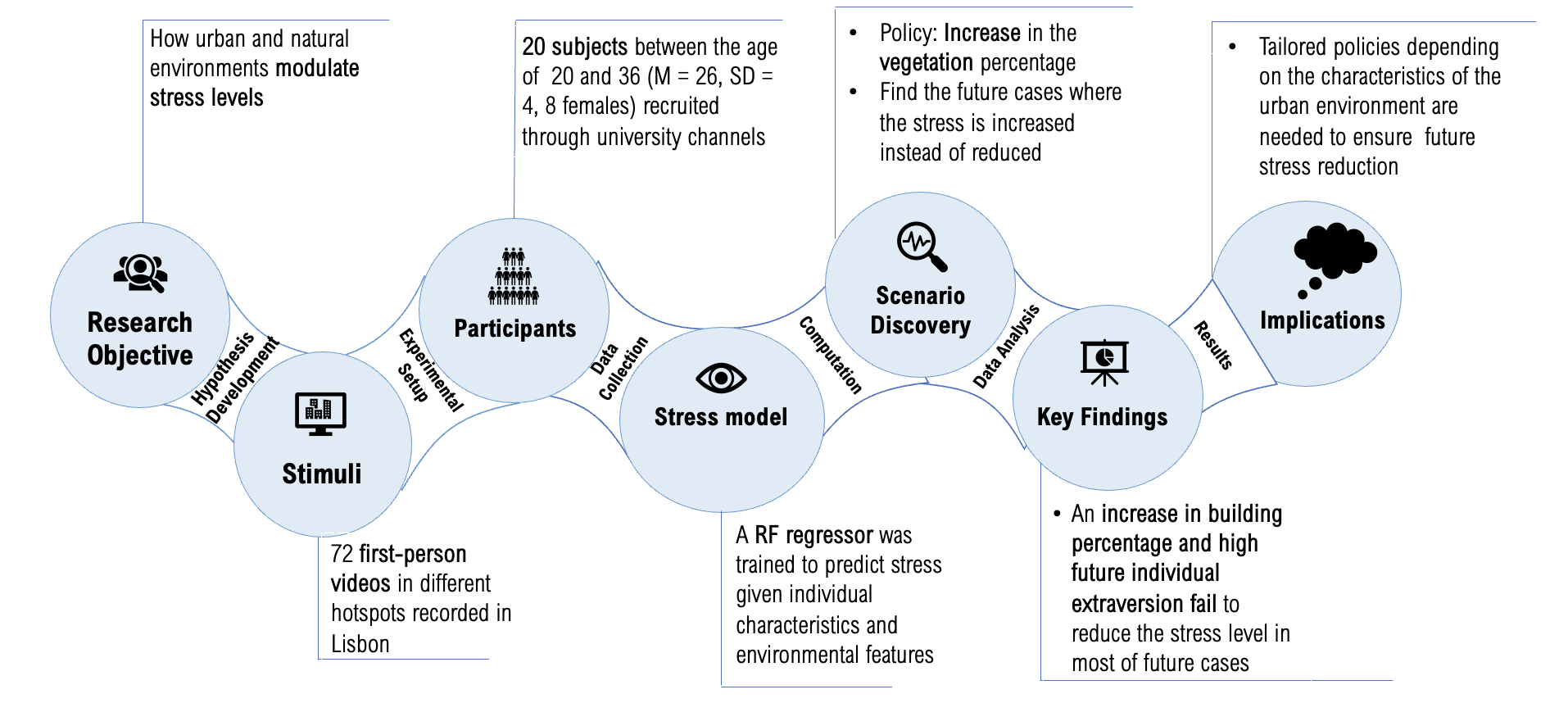}
    \caption{Design overview of the Case Study}
    \label{Case}
\end{figure}

The data from the video-based experiment was used to build a stress model predictor containing individual and environmental features to link the urban spaces with the individual’s well-being. This will give us the Relationships (R) of our XLRM framework from section 3.1.
We then apply this model to predict well-being measures in selected paths in Copenhagen, Denmark, and apply Scenario Discovery to explore the urban environment's role in the stress level under different sources of uncertainty.
Increasing the percentage of vegetation was the policy lever (L) chosen to reduce stress levels, which has multiple documented benefits on mental health outcomes, including reduced stress, anxiety, depression, and aggression \cite{21}. %Then, scenario discovery was applied to investigate the future cases where our selected policy will fail to meet its goal and the conditions needed to reduce stress levels.
Finally, the key findings for each selected path will be explained, and some general implications is presented. An overview of the whole case study setup is illustrated in Figure \ref{Case}.

\subsection{Lisbon data collection}
%As a first step, self-reported evaluation data from Lisbon videos was collected as part of Experiment 2: 'Understanding the neuronal processing of urban space through naturalistic stimuli' of the eMOTIONAL Cities project  \cite{11}.

Firstly, a neuro-urbanism workshop in Lisbon brought together 16 stakeholders from government, academia, non-profits, and urban planning to analyze critical areas using spatial data, provide feedback, and guide neuroscience data collection, resulting in the identification of 24 urban walk paths for data collection \cite{11}.
%a neuro-urbanism workshop was held in Lisbon, bringing together stakeholders and professionals for a participatory analysis aimed at giving feedback on the critical areas identified through a data-driven and spatial data analysis method and guiding the collection of neuroscience data. Sixteen participants participated, including representatives of central and local governments, non-profit organizations, academia, and architectural and urban planning firms. Based on their practical experience and empirical knowledge, this collaborative effort led to the identification of 24 paths where the videos used for this analysis were recorded. \cite{11}.

First-person videos were recorded for the selected paths with a commercial video camera (GoPro Hero 9) at a 1920x1080p resolution by the same individual, who aimed to maintain a consistent walking pace and camera angle (aligned with the direction of movement). %Following the recordings, the clips were edited using Open Shot Editor to optimize scene variety, minimize repetitive scenarios, and remove distracting elements. 
20 adult participants were recruited as a convenient sample for the experiment. %  for the study through the University of Lisbon and the Emotional Cities social media platforms.
None of the participants had a history of psychiatric or neurological disorders. Demographic data (age, gender, and education level) were collected, along with psychological evaluations using the HEXACO Personality Inventory %, a 60-item questionnaire that examines personality traits in six domains, of which Extraversion and Openness to Experience were included in the study
\cite{HEXACO}.

\begin{figure}[H]
    \centering
    \includegraphics[width = 0.7
    \textwidth]{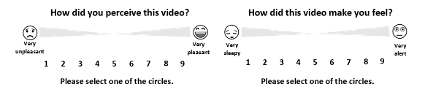}
    \caption{A modified version of the affective slider \cite{12} for valence and arousal evaluation}
    \label{Slider}
\end{figure}

During the experiment, participants viewed the videos from a distance of 50–55 cm on a 21.5” screen with a resolution of 1920x1080. After watching each video, they responded to two affective questions, as illustrated in Figure \ref{Slider}. The first question, 'How did you perceive this video?' was rated on a scale from 1 = very unpleasant to 9 = very pleasant. The second question, 'How did this video make you feel?' was rated from 1 = very sleepy to 9 = very alert.
The final dataset consisted of 1256 observations without missing values from 20 subjects between the ages of  20 and 36 (M = 26, SD = 4, 8 females) recruited through university channels. Participants watched a total of 72 first-person videos of 21 seconds. 

\subsection{Stress model}
\begin{figure}[H]
    \centering
    \begin{subfigure}[b]{0.45\textwidth}
        \centering
        \includegraphics[width=\textwidth]{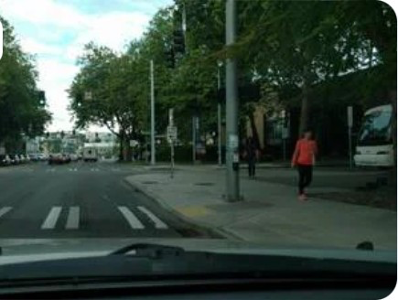} % Cambia "imagen1.jpg" por tu archivo
    \end{subfigure}
    \hfill
    \begin{subfigure}[b]{0.45\textwidth}
        \centering
        \includegraphics[width=\textwidth]{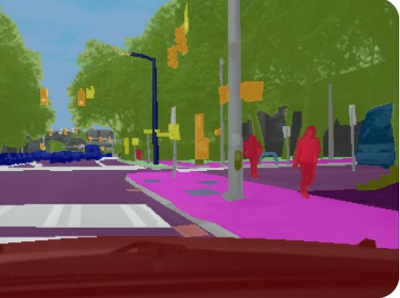} % Cambia "imagen2.jpg" por tu archivo
    \end{subfigure}
    \caption{Example of raw image and annotated image with Mask2Former \cite{17}}
    \label{Mapi}
\end{figure}

The videos from the experiment were then decomposed in frames of frequency 30 FPS, and subsequently, image segmentation of the images was applied to extract valuable information. This step was carried out using Mask2Former, a model that segments images into categories such as roads, buildings, vehicles, and pedestrians \cite{15}, as discussed in the paper "Masked-attention Mask Transformer for Universal Image Segmentation" \cite{16}.  In Figure \ref{Mapi}, we can see an example of a raw image and its annotated image.
After segmenting each image, we calculated the presence of each object type and expressed it as a percentage. We counted the pixels belonging to each category and represented these counts as a percentage of the total number of pixels in the image \cite{17}, \cite{18}.  Finally, the mean value of each feature across all the frames in each path was computed.
Given the process data, we trained a Random Forest Regressor to predict stress levels given some individual characteristics and environmental features. Random forest has multiple known benefits like scalability, robustness to noise and overfitting. Moreover, some studies have shown it as adequate to predict stress levels \cite{RF1}, \cite{RF2}.

\begin{figure}[H]
    \centering
    \begin{subfigure}[b]{0.45\textwidth}
        \centering
        \includegraphics[width=\textwidth]{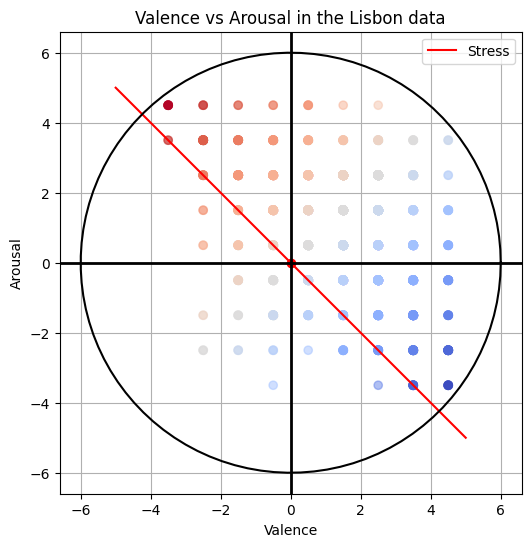}
        \caption{Stress observed in the Lisbon data}
        \label{stress11}
    \end{subfigure}
    \hfill
    \begin{subfigure}[b]{0.45\textwidth}
        \centering
        \includegraphics[width=\textwidth]{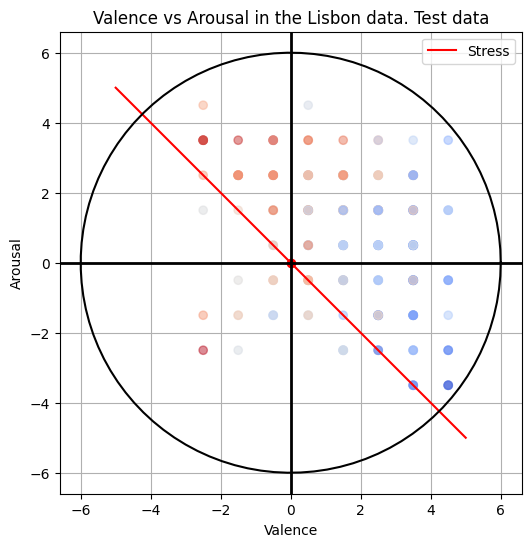}
        \caption{Predicted stress in the Lisbon test data}
        \label{stress22}
    \end{subfigure}
    \caption{Stress representations in the Lisbon data}
    \label{Stress}
\end{figure}

We started by building a model to calculate the projection of the arousal and valence in a 45-degree line, as shown in Figure \ref{stress11}, which, in psychological theory, represents the level of stress \cite{stress}. Blue points, with lower arousal and high valence, represent relaxation. Red points, with high arousal and lower valence, represent stress states.

Cross-validation was used to train the RF Regressor and tune the model hyperparameters to achieve a better fit. The best Random Forest Regressor consisted of 50 decision tree estimators with a maximum depth of 30 each, a minimum sample for leaves of 10 and 2 minimum samples to perform a split. Moreover, bootstrap %(a sampling technique that involves repeatedly drawing random samples with replacement from a dataset)
was employed in the estimation process.\\
The coefficient of determination ($R^2$) was used in training to fit the model to the data. $R^2$  indicates the proportion of the variance in the dependent variable that is predictable from the independent variables,
\begin{equation}
R^2 = 1 - \frac{\sum_{i=1}^{n} (y_i - \hat{y}_i)^2}{\sum_{i=1}^{n} (y_i - \bar{y})^2}
\label{eq:1}
\end{equation}
where $y_i$ represents the actual value, $\hat{y}_i$  represents the predicted value, $\bar{y}$ is the mean of the actual values, and $n$ is the number of observations in the test set. Therefore, the closer the $R^2$  is to 1, the greater the variance explained and the better the fit of the model. Even though the valence and arousal values are discrete, the combination of them in the stress variable make it continuous, and its prediction works better with regression models.\\
In our case $R^2=0.53$,  MSE (Mean Absolute Error) = 2.05 and MAE ((Mean Squared Error) = 1.14,  which indicates that there is still some data variability that is not explained by the model structure. This could be due to the small sample size of the data and the lack of recorded variables that influence the stress response.\\
Moreover, to understand how well our proposed model predicts the stress level, we plot the stress prediction for the test data given the valence and arousal coordinates, as shown in Figure \ref{stress22}. The test data was 20\% of the total data randomly selected and as it can be seen in Figure \ref{stress22}, even though there are some misleading points, i.e., some slightly blue points in the upper left corner and some slightly red points in the lower right corner, the model seems to capture the overall stress trend in the test data. The range of the predicted stress goes from [-4, 3.2] where positive values of stress represent higher levels and negative values refer to calmer states.

\begin{figure}[H]
    \centering
    \includegraphics[width = 0.6
    \textwidth]{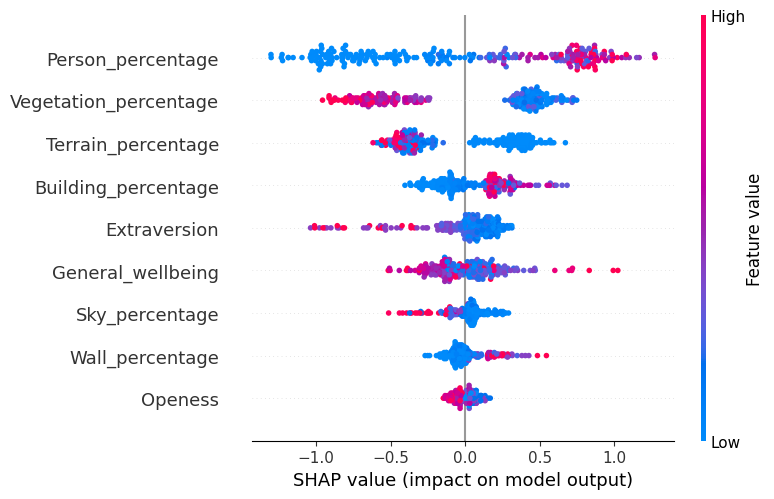}
    \caption{Proposed Scenario Discovery framework}
    \label{shap}
\end{figure}

Finally, the SHAP values were calculated to interpret the role of each input variable played in predicting the stress level. SHAP values (Shapley Additive exPlanations) quantify the contribution of each feature to the prediction by distributing the difference between the model's output and a baseline value among the input features. As shown in Figure \ref{shap}, the higher the percentage of people and buildings, the higher the predicted stress level, consistent with earlier psychological findings \cite{people} \cite{urban}. The same theoretical correspondence happens with the vegetation percentage, where the higher the value, the lower the predicted stress. Access to green spaces has been associated with improved mental health outcomes, including decreased stress, anxiety, depression, and aggression \cite{21}. Regarding the observer characteristics, openness and extraversion, both personality traits described in the HEXACO Personality Inventory \cite{HEXACO}, show an impact on stress outcomes. Personality traits like openness to experience, extraversion, and neuroticism have previously been linked to emotional responses, behaviour, and cognition \cite{personality1}). For example, studies have shown that extroverts respond more strongly to positive affect than introverts, reflecting distinct patterns of emotional engagement with their surroundings \cite{personality2}. Similarly, openness to experience, marked by curiosity and a willingness to engage with new sensations, has been closely linked with Connectedness to Nature, or one’s sense of belonging to the natural world \cite{personality3}.

% Personality traits significantly influence how people perceive, react to and cope with stress \cite{14}. Personality helps explain why people think, feel, and behave differently or similarly in various situations. Evidence supports that personality factors, more so than the environment, play a causal role in how individuals cope and their stress reactions \cite{13}. Greater openness mainly results in beneficial effects on the stress process. Similarly, extroverted individuals generally handle stress better, largely due to their increased sociability. This sociability enables extraverted individuals to seek and receive more social support \cite{13}.

Regarding the stress predictor model results, to have an order of magnitude of some of the more important features, the mean value through the 72 Lisbon videos of some of the input variables is presented in Table 1. As we can see in the table, the value of the features varies considerably from video to video since they are expected to capture a wide variety of urban sceneries.
\begin{table}[H]
\centering
\resizebox{\textwidth}{!}{
\begin{tabular}{|l|c|c|c|}
\hline
\textbf{Features} & \textbf{Mean value over all the videos} & \textbf{Minimum mean value over a video} & \textbf{Maximum mean value over a video} \\ \hline
Building percentage (\%) & 26.47 & 0.0 & 96.07 \\ \hline
Vegetation percentage (\%) & 33.31 & 0.07 & 95.75 \\ \hline
Person percentage (\%) & 10.23 & 0.0 & 46.12 \\ \hline
Extraversion & 3.12 & 3.02 & 3.14 \\ \hline
\end{tabular}
}
\caption{Summary of feature statistics across videos.}
\label{tab:feature_stats}
\end{table}

\subsection{Copenhagen data collection}
Given the stress level predictor model trained with the Lisbon path data, we extrapolate these results and apply them to different paths in Greater Copenhagen to better understand how different urban and individual features affect stress prediction. We aim to develop a policy that will reduce stress levels in Copenhagen in a wide variety of future scenarios.

First, we selected four paths throughout Copenhagen, each representing different city areas with distinct environmental features. The selection depended on a hot-spot spatial analysis carried out on spatially aggregated data on demographics, socio-economic, land-use and urban features,  mobility, pollution, noise and its relations with both mental and physical health records. We refer to \cite{20} for a detailed technical description of the methods in this path selection, but we reiterate that the four selected paths are herein used for scenario discovery showcasing purposes. However, the existing heterogeneity in these four paths already enables us to explore how policies evolve in different environments.

Next, we extracted the relevant model input data for each path from Mapillary street images \footnote{Mapillary is a collaborative platform that enables users to capture, share, and explore street-level imagery from around the world. \textit{https://www.mapillary.com}}.For a sample of points in the path, we collected all Mapillary images in 50x50 meter cells, whose center was defined as the sampled coordinate. Consequently, the number of images retrieved per coordinate varied. Here, we assumed that Mapillary images work as a proxy for the person walks videos collected in Lisbon since no videos were recorded in Copenhagen.

Subsequently, we applied image segmentation to the Mapillary images to extract valuable information. This step was also carried out using Mask2Former. Once all individual images were processed, we averaged the features to compute the mean percentage for each object type across all images at the same coordinate. This averaging reveals the most prevalent features in images from specific geographic locations, offering an average representation of the scenery surrounding a given coordinate \cite{19}.

\begin{figure}[H]
    \centering
    % Primera fila
    \begin{subfigure}[b]{0.4\textwidth} % Ajusta el ancho según sea necesario
        \centering
        \includegraphics[width=0.75\textwidth]{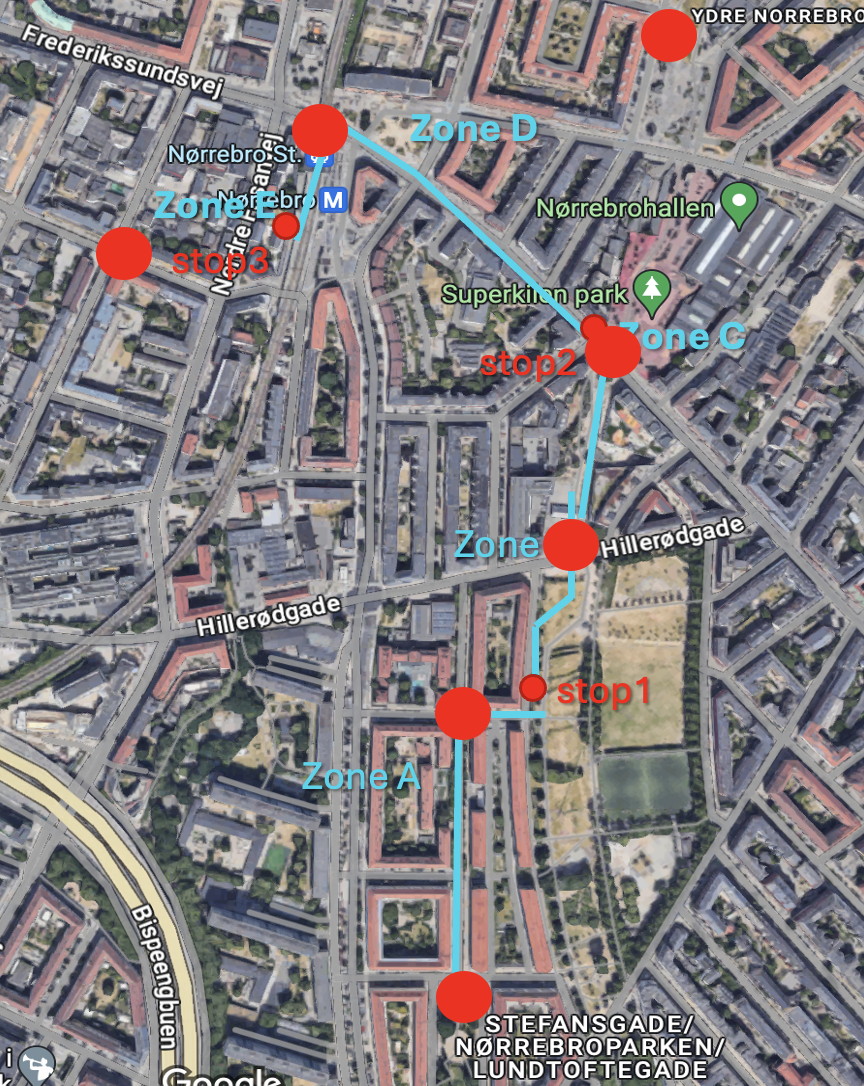} % Cambia "imagen1.jpg" por tu archivo
        \caption{Collected data points in the Nørrebro path}
    \end{subfigure}
    \hspace{1em} % Espacio horizontal entre imágenes
    \begin{subfigure}[b]{0.4\textwidth}
        \centering
        \includegraphics[width=\textwidth]{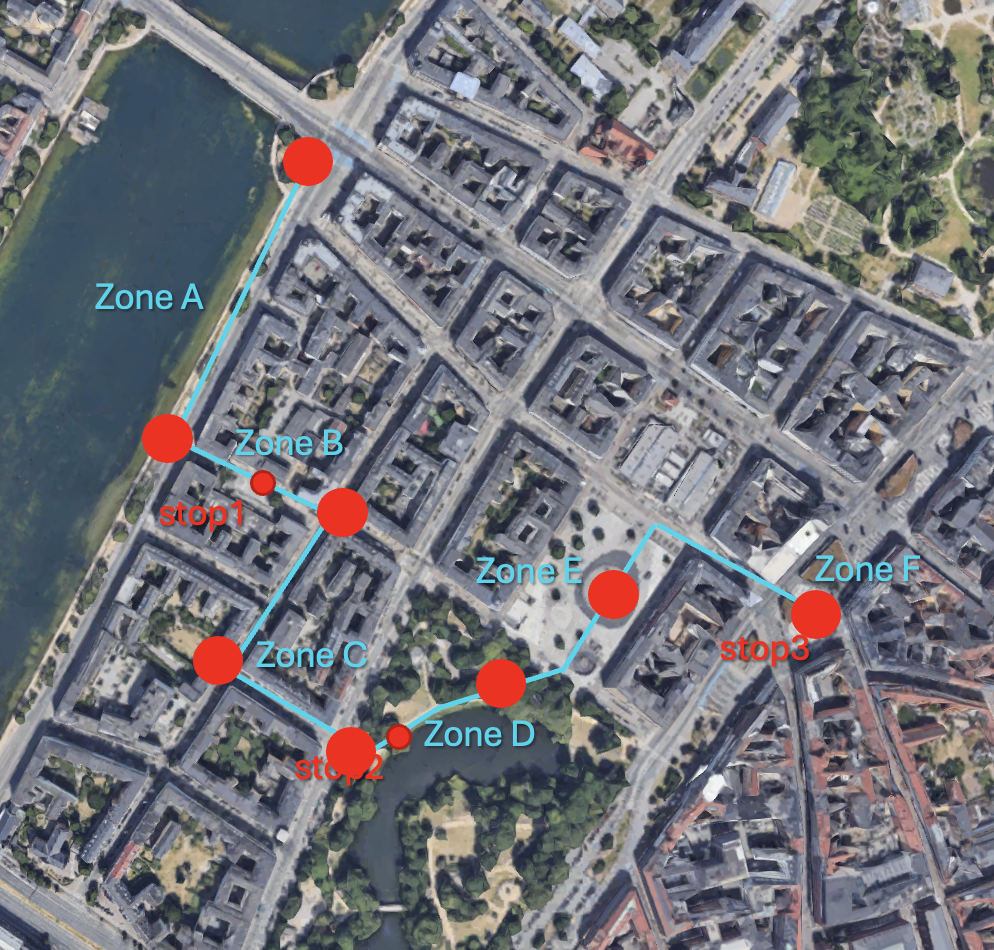} % Cambia "imagen2.jpg" por tu archivo
        \caption{Collected data points in the Nørreport path}
    \end{subfigure}
    % Espacio entre filas
    \vspace{1em}
    % Segunda fila
    \begin{subfigure}[b]{0.4\textwidth}
        \centering
        \includegraphics[width=\textwidth]{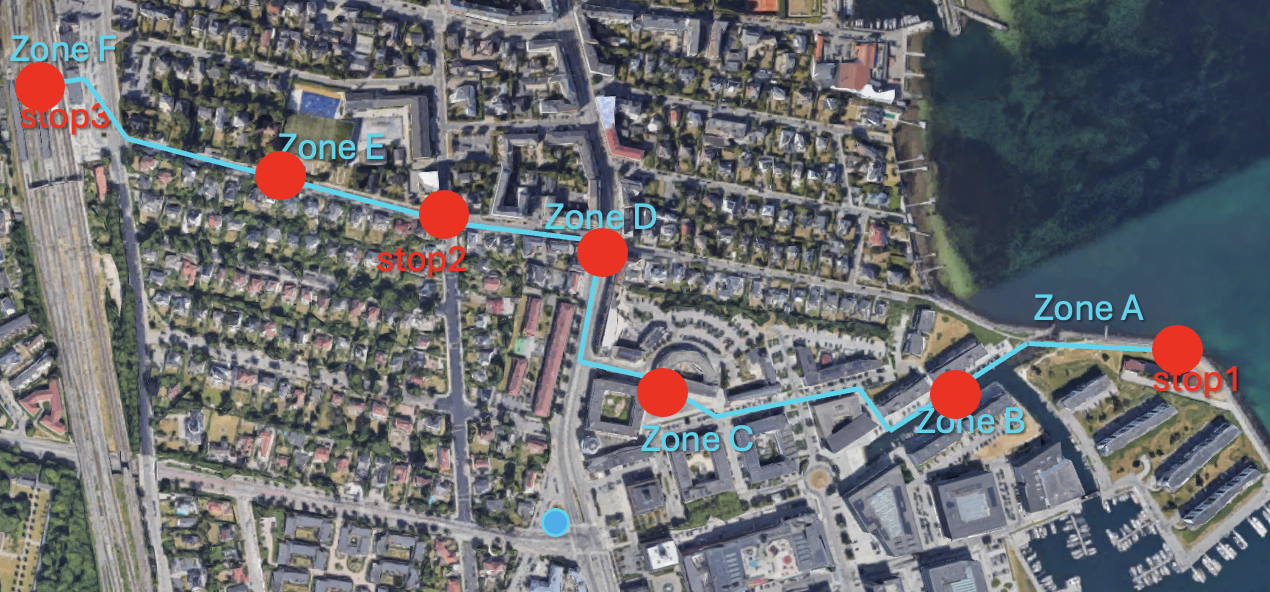} % Cambia "imagen3.jpg" por tu archivo
        \caption{Collected data points in the Hellerup path}
        \label{fig:imagen3}
    \end{subfigure}
    \hspace{1em} % Espacio horizontal entre imágenes
    \begin{subfigure}[b]{0.4\textwidth}
        \centering
        \includegraphics[width=\textwidth]{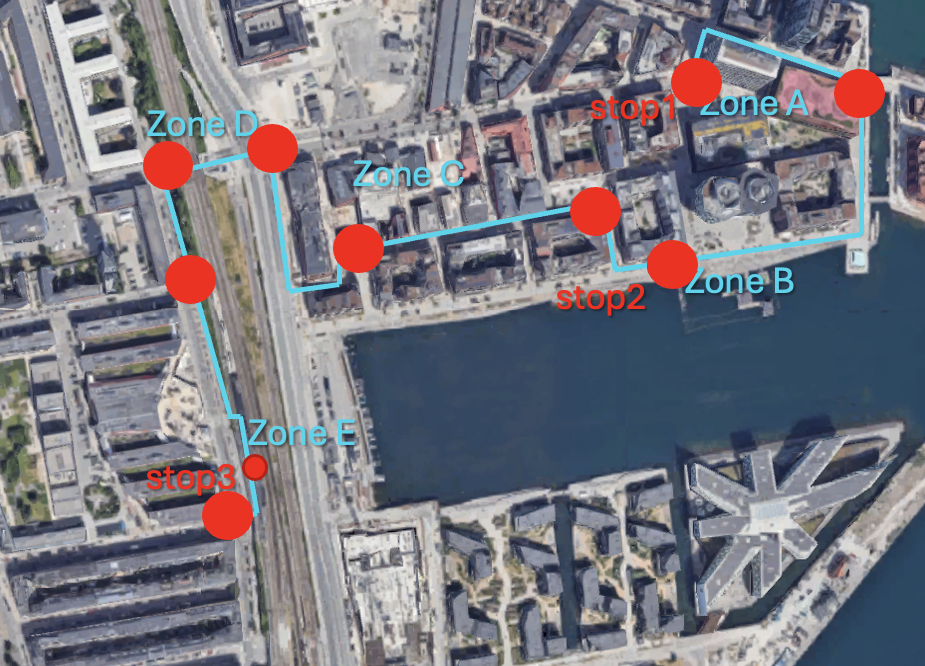} % Cambia "imagen4.jpg" por tu archivo
        \caption{Collected data points in the Nordhavn path}
        \label{1}
    \end{subfigure}
    
    \caption{Collected data points in the Greater Copenhagen area}
    \label{cph}
\end{figure}

% \begin{comment}
% \begin{figure}[H]
%     \centering
%     \includegraphics[width = 0.85
%     \textwidth]{sources/Imagenes/Cph.png}
%     \caption{Hotspots of mental and physical health in the Greater Copenhagen area and the location of the selected paths \cite{20}}
%     \label{Hotspot}
% \end{figure}
% \end{comment}
In Figure \ref{cph}, we plot the selected paths with the data points where the feature percentages from the Mapillary images were extracted. The sampled coordinates were selected based on the paths used for Experiment 4: Outdoor neuroscience experiment: “eMOTIONAL cities walker” conducted in Copenhagen by the eMOTIONAL Cities project, the location of the point within the paths was such that it captures the maximum variability of urban elements and expected emotional states.

\begin{figure}[H]
    \centering
    \begin{subfigure}[b]{0.45\textwidth}
        \centering
        \includegraphics[width=\textwidth]{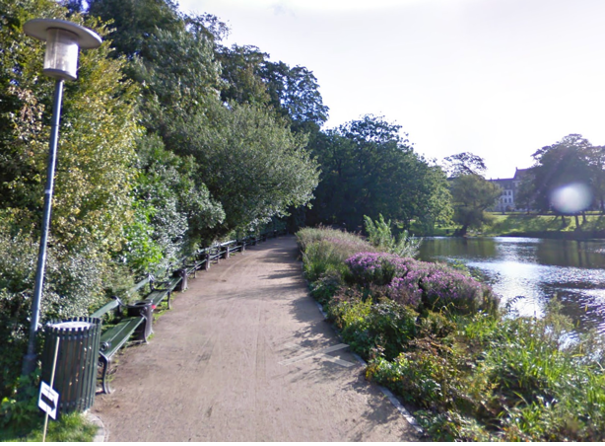} % Cambia "imagen1.jpg" por tu archivo
    \end{subfigure}
    \hfill
    \begin{subfigure}[b]{0.45\textwidth}
        \centering
        \includegraphics[width=\textwidth]{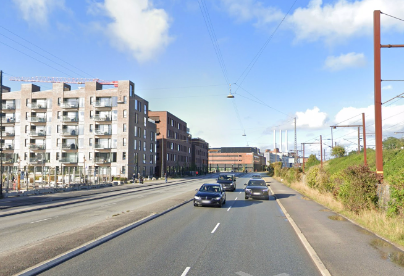} % Cambia "imagen2.jpg" por tu archivo
    \end{subfigure}
    \caption{Google Images taken form the Nørreport and Norhavn paths respectively}
    \label{Path}
\end{figure}

Then, the average across points within each path was calculated, and some of the most important features for Scenario Discovery are presented in Table \ref{table_cph}. Notably, Nørreport has the highest percentage of vegetation, 16.57 \%, while %due to its proximity to the lake parks and Ørstedsparken garden. 
 Nordhavn has the lower, 6.25\%, as shown in Figure \ref{Path}. %, since it is a newly urbanized area located near train rails.
It is important to highlight that the percentage of people captured in the images is significantly low compared with the data from Lisbon (Table 1) since Mapillary images are often collected on the road pavement, instead of the sidewalk. However, this does not imply that these areas are not densely populated. To address this bias, we adjusted the person percentage upwards by 10\%, the mean of the Lisbon data, before normalizing it with the other features for all images from Copenhagen.

\begin{table}[H]
    \centering
    \begin{tabular}{|>{\centering\arraybackslash}m{5cm}|c|c|c|c|}
        \hline
        \textbf{Features} & \textbf{Nørrebro} & \textbf{Nørreport} & \textbf{Hellerup} & \textbf{Nordhavn} \\ \hline
        Vegetation percentage (\%) & 10.22 & 16.57 & 12.6 & 6.25 \\ \hline
        Building percentage (\%) & 24.7 & 22 & 25 & 26.5 \\ \hline
        Person percentage (\%) & 0.82 & 0.46 & 0.24 & 0.26 \\ \hline
    \end{tabular}
    \caption{Percentage of environmental features for each considered Copenhagen path.}
    \label{table_cph}
\end{table}

\subsection{Scenario Discovery framework for the Case Study}
Based on the information from our stress predictor model and the characteristics of the Copenhagen paths, we examined which policies could be applied in this case study to reduce stress levels.

We chose to ensure better access to vegetation spaces and nature, as it has been associated with improved mental health outcomes, including decreased stress, anxiety, depression, and aggression. This is particularly important for individuals living in densely populated areas who lack personal yard spaces. Additionally, high noise levels in the city center can disrupt sleep and relaxation, and research has demonstrated that planting trees can help reduce noise levels. Lastly, air pollution is linked to poorer mental health outcomes, and increasing vegetation can mitigate its impact \cite{21}. %Other dimensions of interest in this case study are blue spaces from the lakes and the canals around the city and walkability and mobility, especially by bike, since Copenhagen is well known for its high bike use.
Table \ref{SD} shows the XLRM Framework proposed for this case study where a vulnerable future case is defined as a scenario where the stress is not reduced but increased after the policy application.

\begin{table}[H]
    \centering
    \begin{tabular}{|p{6cm}|p{6cm}|}
        \hline
        \textbf{Uncertainties (X)} &         \textbf{ Policy levers (L)} \\ \hline
        Individual characteristics: extraversion \newline Environmental features: building percentage and person percentage & Increase the vegetation percentage in each point of the selected path \\ \hline
        \textbf{Relationships (R)} & \textbf{Measures (M)} \\ \hline
        The results from the stress model applied in different paths in Copenhagen & Reducing the number of vulnerable cases for various future scenarios. \\ \hline
    \end{tabular}
    \caption{XLRM Framework for our Case Study}
    \label{SD}
\end{table}

We focus on three key uncertainties in our Scenario Discovery Framework. Two of these are environmental factors with significant influence on the stress predictor model: building percentage and person percentage. We chose them because of their high impact on predicting the stress response according to the SHAP values. Moreover, building and person percentages are increasing nowadays in all cities and are expected to continue increasing in the future. So, high variability of these features is expected. For both, we define a variability range from 50\% to 150\% of their baseline values. This range will be later adjusted due to the normalization process, which ensures that the combined percentage of all feature types at any coordinate along the path does not exceed 100\%. Therefore, the 50\% and 100\% values were chosen to physically represent a percentage variation of the current values within some control limits.

The third uncertainty is related to the observer and is extraversion \cite{HEXACO}. We chose extraversion because it was the individual characteristic that most impacted the stress predictor model, and we consider it helpful to have an uncertainty variable at the individual’s level to see its impact on the predicted results. In the Lisbon dataset, extraversion has a mean value of 3.12, with a minimum of 2.41 and a maximum of 5. Since the Copenhagen data lack personality trait information, the Lisbon average is used as the baseline scenario. For uncertainty exploration, a range between 80\% and 140\% of this value is applied, also in concordance with the extraversion reported in Lisbon. Lower levels of extraversion refer to somewhat introverted individuals who prefer more limited interactions,  while higher levels represent highly sociable and enthusiastic people. Table \ref{tab:uncertainty} summarizes the uncertainty selection and their range, by stress-testing urban policies against hypothetical disruptions, we can identify vulnerabilities and design adaptive governance frameworks that enhance urban resilience \cite{zenia4}
\begin{table}[H]
    \centering
    \begin{tabular}{|l|c|c|}
        \hline
        \textbf{Uncertainty} & \textbf{Minimum value} & \textbf{Maximum value} \\ \hline
        Building percentage & 50\% of its value & 150\% of its value \\ \hline
        Person percentage & 50\% of its value & 150\% of its value \\ \hline
        Extraversion & 80\% of 3.12 = 2.5 & 140\% of 3.12 = 4.37 \\ \hline
    \end{tabular}
    \caption{Selected uncertainty features and their ranges.}
    \label{tab:uncertainty}
\end{table}
\section{Results}

\subsection{Percentage of vegetation needed}
Firstly, we investigate the percentage of vegetation required to reduce stress levels. We run 200 and 300 future LHS scenarios for various vegetation percentages. For three uncertainty variables, 200 runs give a relative density metric, $J$, of 5.8, which is considered appropriate given the standards.

\begin{figure}[H]
    \centering
    % Primera fila
    \begin{subfigure}[b]{0.4\textwidth} % Ajusta el ancho según sea necesario
        \centering
        \includegraphics[width=\textwidth]{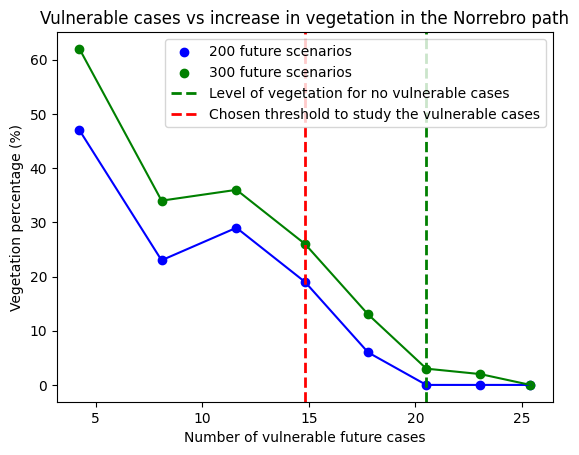} % Cambia "imagen1.jpg" por tu archivo
        \caption{Percentage of vegetation for the Nørrebro path}
    \end{subfigure}
    \hspace{1em} % Espacio horizontal entre imágenes
    \begin{subfigure}[b]{0.4\textwidth}
        \centering
        \includegraphics[width=\textwidth]{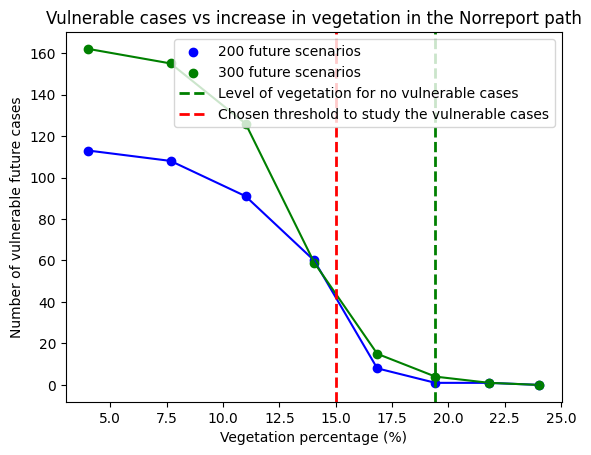} % Cambia "imagen2.jpg" por tu archivo
        \caption{Percentage of vegetation for the Nørreport path}
    \end{subfigure}
    % Espacio entre filas
    \vspace{1em}
    % Segunda fila
    \begin{subfigure}[b]{0.4\textwidth}
        \centering
        \includegraphics[width=\textwidth]{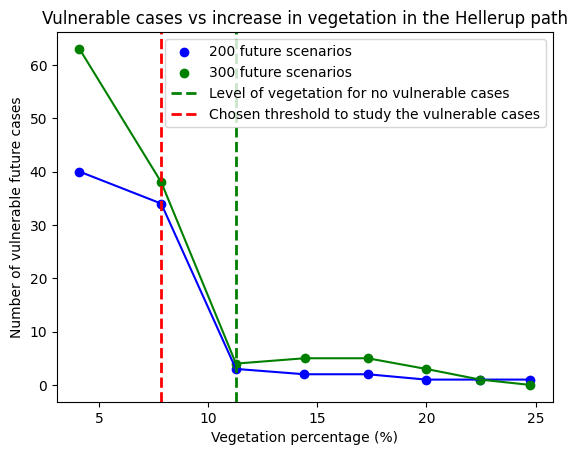} % Cambia "imagen3.jpg" por tu archivo
        \caption{Percentage of vegetation for the Hellerup path}
    \end{subfigure}
    \hspace{1em} % Espacio horizontal entre imágenes
    \begin{subfigure}[b]{0.4\textwidth}
        \centering
        \includegraphics[width=\textwidth]{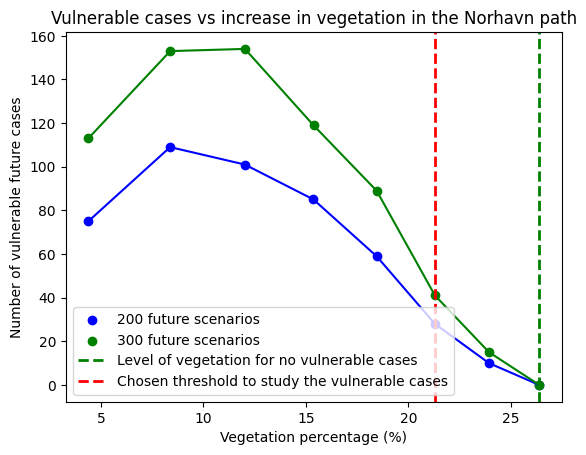} % Cambia "imagen4.jpg" por tu archivo
        \caption{Percentage of vegetation for the Nordhavn path}
    \end{subfigure}
    
    \caption{Percentage of vegetation needed for each of the paths}
    \label{vege}
\end{figure}

Figure \ref{vege} illustrates the number of vulnerable cases for each combination of vegetation percentage across 200 and 300 LHS scenarios for each selected path. Two key insights can be drawn from this figure. The first one is that increasing the number of scenarios to 300 does not change the vegetation percentage threshold at which the number of vulnerable cases is zero or nearly zero. Consequently, there is no additional benefit to running 300 model simulations, and we will proceed with our analysis using 200 LHS future cases. It is important to note that this comparison was possible due to the short estimation time of the model. Running a larger number of simulations with a more complex and time-consuming model would have been unfeasible and, therefore, we would need to resort to the active learning metamodeling approach explained in Section 3.

The second insight from Figure \ref{vege} is that the percentage of vegetation required to achieve our policy goals varies significantly depending on the characteristics of each path. Specifically, approximately 15\% of vegetation is required in Norrrebro and Nørreport, while only 7\% is needed in Hellerup, and up to 22\% in Nordhvan. This is a highly valuable insight from an urban planning perspective as it allows the creation of more tailored policies that would work in various future scenarios while reducing the resources needed in some areas and maximizing the efficiency of the chosen policies.

\subsection{Scenario Discovery results for each path}
For each of the selected paths, Scenario Discovery is performed with a previously selected vegetation threshold to study the location of the vulnerable cases, i.e. the futures that experience an increase instead of a reduction of the stress levels, on average for all the points in each path, when comparing with the baseline case (no policy applied). \\
Identifying the location of vulnerable cases provides policymakers and stakeholders with a better understanding of the conditions under which the proposed policy will not perform as expected. This information can help monitor certain future uncertainties to be prepared for when the proposed policy does not serve its purpose and needs to be replaced.

\subsection{Nørrebro path results}
\begin{figure}[H]
    \centering
    \begin{subfigure}[b]{0.434\textwidth}
        \centering
        \includegraphics[width=\textwidth]{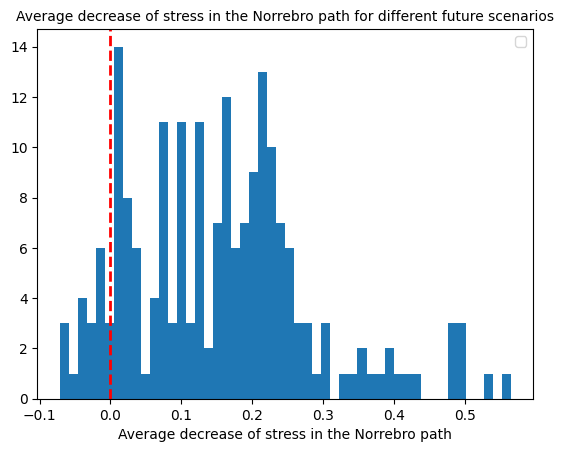} % Cambia "imagen1.jpg" por tu archivo
        \caption{Stress reduction}
        \label{Norrebro2}
    \end{subfigure}
    \hfill
    \begin{subfigure}[b]{0.45\textwidth}
        \centering
        \includegraphics[width=\textwidth]{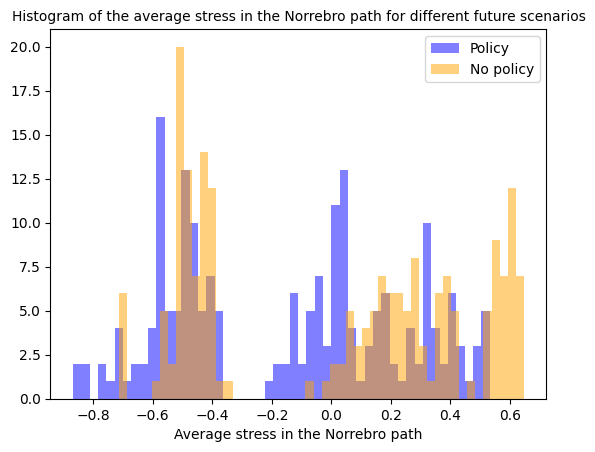} % Cambia "imagen2.jpg" por tu archivo
        \caption{Stress levels before and after the policy}
        \label{Norrebro3}
    \end{subfigure}
    \caption{Stress before and after applying the policy in the Nørrebro path}
    \label{hola}
\end{figure}

After normalisation and scenario generation, the average increase of vegetation in Nørrebro is 14.81\% and with this percentage, the number of vulnerable cases is 19.
Figure \ref{Norrebro2} shows the decrease of stress compared to the baseline case average in the points in the Nørrebro path for the 200 considered future scenarios. As we showed in Section 5.1, most cases experience a stress reduction for the selected percentage of vegetation, with the maximum decrease around 0,6 given the scale presented in Figure \ref{Stress}. On Figure \ref{Norrebro3}, the average stress is computed with and without applying the policy for the same 200 LHS point.

\begin{figure}[H]
    \centering
    \includegraphics[width = 0.4
    \textwidth]{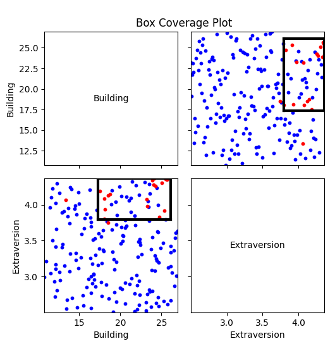}
    \caption{Found PRIM box for the vulnerable cases in the Nørrebro path}
    \label{PRIM1}
\end{figure}

To highlight the location of the vulnerable cases, PRIM algorithm was employed.
Figure \ref{PRIM1} shows the PRIM box where most vulnerable cases are concentrated (coverage = 89,5\%). Notice that it corresponds to scenarios where the building percentage goes above 17,3\% and the extraversion is more than 3,80. \\
Although we previously noted that high extraversion acts as a protective factor against stress, the stress reduction compared to the baseline scenario is less pronounced in individuals with higher extraversion than in those who are less extroverted. 
Intuitively, this means that even though the stress levels could be lower for extroverted individuals, the reduction in stress is smaller for them when planting more trees. This effect coincides with an increase in building percentage. As the building percentage rises, available space for vegetation decreases, leading to higher predicted stress levels. Moreover, the stress predictor model indicates a strong correlation between high building percentages and elevated stress levels.

\subsection{Nørreport path results}
\begin{figure}[H]
    \centering
    \begin{subfigure}[b]{0.434\textwidth}
        \centering
        \includegraphics[width=\textwidth]{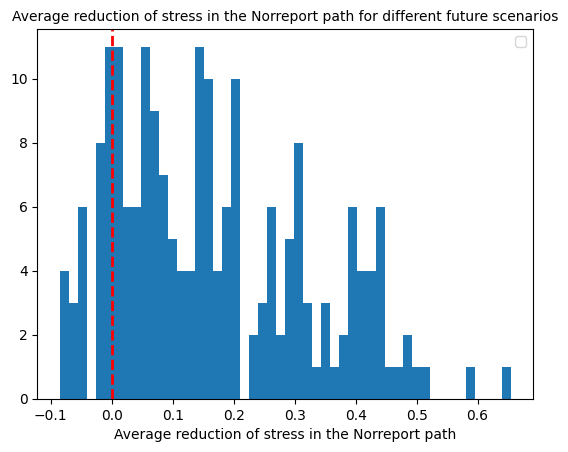} % Cambia "imagen1.jpg" por tu archivo
        \caption{Stress reduction}
        \label{Norreport1}
    \end{subfigure}
    \hfill
    \begin{subfigure}[b]{0.45\textwidth}
        \centering
        \includegraphics[width=\textwidth]{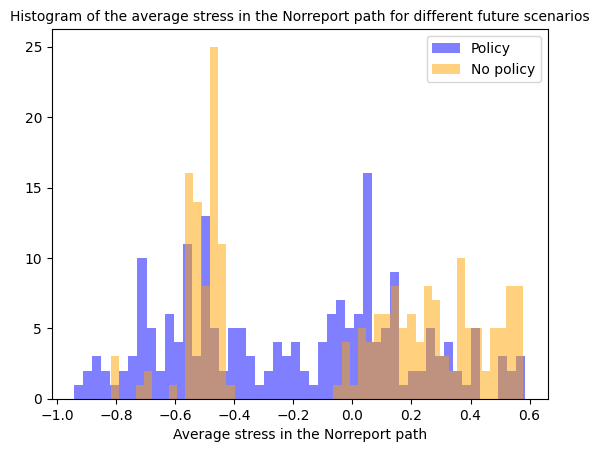} % Cambia "imagen2.jpg" por tu archivo
        \caption{Stress levels before and after the policy}
        \label{Norreport2}
    \end{subfigure}
    \caption{Stress before and after applying the policy in the Nørreport path}
\end{figure}

After normalisation and scenario generation, the average increase of vegetation in Nørreport is 15.18\%, and with this percentage, the number of vulnerable cases is 29.
Figure \ref{Norreport1} shows the decrease of stress compared to the baseline case average in the points in the Nørreport path for the 200 considered future scenarios, where most cases experience a stress reduction for the selected percentage of vegetation with the maximum decrease around 0,7. On Figure \ref{Norreport2}, the average stress is computed with and without applying the policy for the same 200 LHS point.

\begin{figure}[H]
    \centering
    \includegraphics[width = 0.45
    \textwidth]{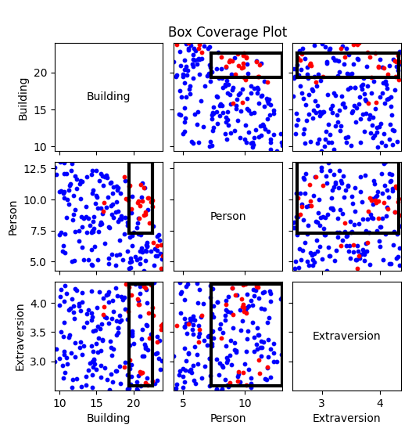}
    \caption{Found PRIM box for the vulnerable cases in the Nørreport path}
    \label{PRIM2}
\end{figure}

Figure \ref{PRIM2} presents the identified PRIM box for vulnerable cases along the Nørreport path. Here, a higher building percentage is associated with an increase in vulnerable cases, especially when the percentage of people exceeds 7\%. However, unlike other paths, extraversion does not appear to be a key factor in determining the locations of these vulnerable cases.

\subsection{Hellerup path results}
\begin{figure}[H]
    \centering
    \begin{subfigure}[b]{0.434\textwidth}
        \centering
        \includegraphics[width=\textwidth]{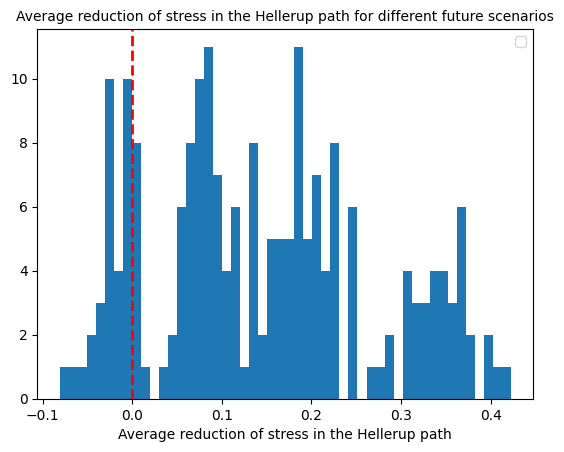} % Cambia "imagen1.jpg" por tu archivo
        \caption{Stress reduction}
        \label{hellerup1}
    \end{subfigure}
    \hfill
    \begin{subfigure}[b]{0.45\textwidth}
        \centering
        \includegraphics[width=\textwidth]{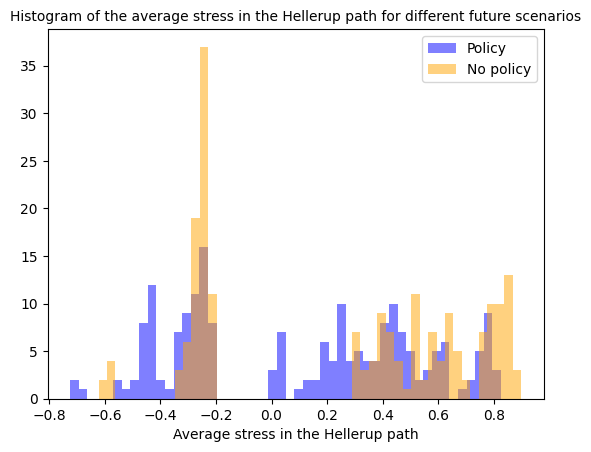} % Cambia "imagen2.jpg" por tu archivo
        \caption{Stress levels before and after the policy}
        \label{hellerup2}
    \end{subfigure}
    \caption{Stress before and after applying the policy in the Hellerup path}
\end{figure}

After normalisation and scenario generation, the average increase of vegetation in Hellerup is 7.86\%, and with this percentage, the number of vulnerable cases is 34.
Figure \ref{hellerup1} shows the decrease of stress compared to the baseline case average in the points in the Hellerup path for the 200 considered future scenarios where most cases experience a stress reduction for the selected percentage of vegetation,  with the maximum decrease around 0,4. On Figure \ref{hellerup2}, the average stress is computed with and without applying the policy for the same 200 LHS point. 

\begin{figure}[H]
    \centering
    \includegraphics[width = 0.45
    \textwidth]{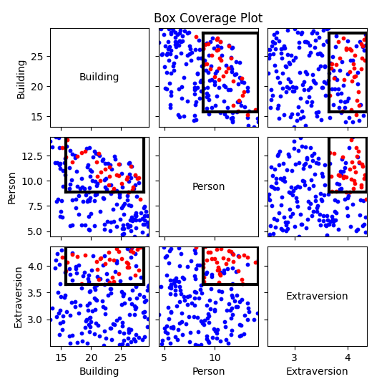}
    \caption{Found PRIM box for the vulnerable cases in the Hellerup path}
    \label{PRIM3}
\end{figure}

Figure \ref{PRIM3} displays the identified PRIM box containing vulnerable cases along the Hellerup path. Here, an increase in people and extraversion appears to be the primary drivers of the prevalence of vulnerable cases. In contrast, the building percentage seems to have less influence, as its range is broader in this scenario.

\subsection{Nordhavn path results}
\begin{figure}[H]
    \centering
    \begin{subfigure}[b]{0.434\textwidth}
        \centering
        \includegraphics[width=\textwidth]{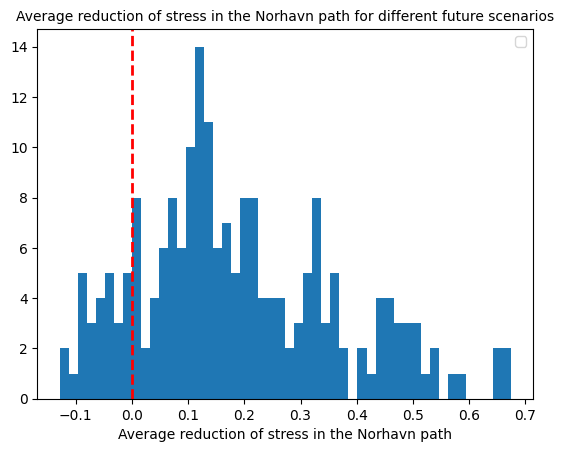} % Cambia "imagen1.jpg" por tu archivo
        \caption{Stress reduction}
        \label{nordhavn1}
    \end{subfigure}
    \hfill
    \begin{subfigure}[b]{0.45\textwidth}
        \centering
        \includegraphics[width=\textwidth]{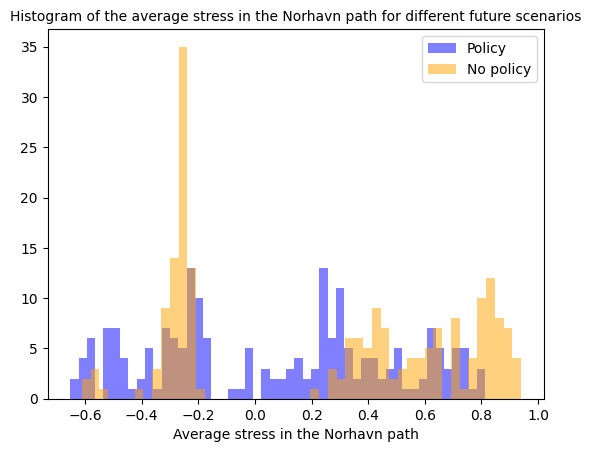} % Cambia "imagen2.jpg" por tu archivo
        \caption{Stress levels before and after the policy}
        \label{nordhavn2}
    \end{subfigure}
    \caption{Stress before and after applying the policy in the Nordhavn path}
\end{figure}

\begin{figure}[H]
    \centering
    \includegraphics[width = 0.45
    \textwidth]{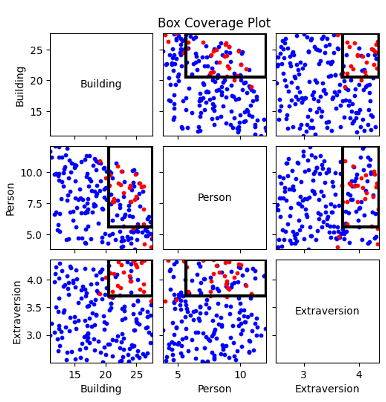}
    \caption{Found PRIM box for the vulnerable cases in the Nordhavn path}
    \label{PRIM4}
\end{figure}

After normalisation and scenario generation, the average increase of vegetation in Nordhavn is 28\%, and with this percentage, the number of vulnerable cases is 28. As mentioned before, this path had the lowest level of vegetation and is close to the train lines, which aligns with a higher need for vegetation.
Figure \ref{nordhavn1} shows the decrease of stress compared to the baseline case average in the points in the Nordhavn path for the 200 considered future scenarios. Most cases experience a stress reduction for the selected percentage of vegetation, with the maximum decrease around 0,7. On Figure \ref{nordhavn2} the average stress is computed with and without applying the policy for the same 200 LHS point.
Figure \ref{PRIM4} illustrates the identified PRIM box containing vulnerable cases. Similar to the Nørrebro case, the likelihood of encountering a vulnerable scenario increases with higher percentages of buildings and extraversion. While the percentage of people appears to contribute somewhat to the rise in vulnerable cases, the correlation is weak due to its wide range of variability.

\subsection{Results of active learning sampling strategies}
\begin{figure}[H]
    \centering
    \begin{subfigure}[b]{0.434\textwidth}
        \centering
        \includegraphics[width=\textwidth]{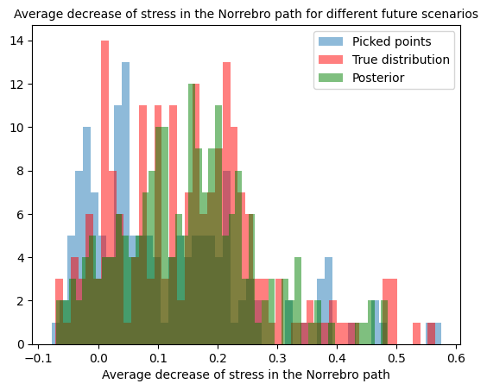} % Cambia "imagen1.jpg" por tu archivo
        \caption{Distribution of the reduction of stress in Nørrebro for the picked points, the true distribution and the posterior distribution}
        \label{AL1}
    \end{subfigure}
    \hfill
    \begin{subfigure}[b]{0.45\textwidth}
        \centering
        \includegraphics[width=\textwidth]{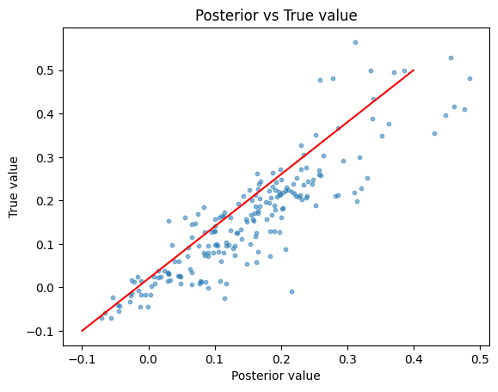} % Cambia "imagen2.jpg" por tu archivo
        \caption{Correlation between the true values and the posterior distribution}
         \label{AL2}
    \end{subfigure}
    \caption{Results of Algorithm 1}
\end{figure}

\begin{figure}[H]
    \centering
    \begin{subfigure}[b]{0.434\textwidth}
        \centering
        \includegraphics[width=\textwidth]{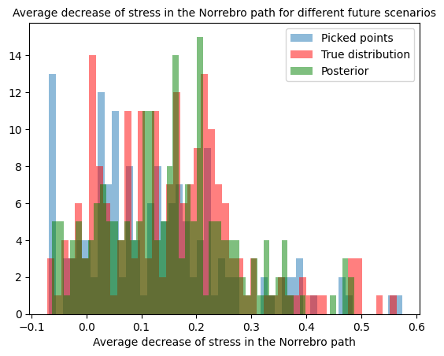} % Cambia "imagen1.jpg" por tu archivo
        \caption{Distribution of the reduction of stress in Nørrebro for the picked points, the true distribution and the posterior distribution}
        \label{AL3}
    \end{subfigure}
    \hfill
    \begin{subfigure}[b]{0.45\textwidth}
        \centering
        \includegraphics[width=\textwidth]{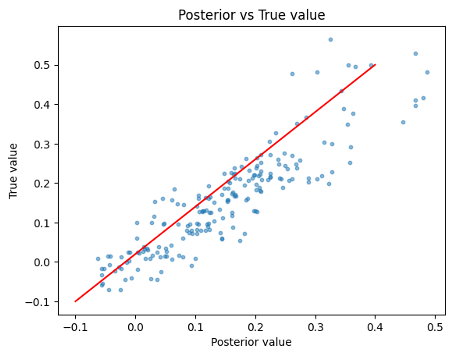} % Cambia "imagen2.jpg" por tu archivo
        \caption{Correlation between the true values and the posterior distribution}
        \label{AL4}
    \end{subfigure}
    \caption{Results of Algorithm 2}
\end{figure}

In this last section of the results, PRIM is combined with Active Learning to reduce the number of sample points needed to arrive at a similar conclusion.
The Nørrebro path was selected for this proof-of-concept, and 100 LHS initial sampled points were run with the stress model. After that, Algorithm 1 was employed to sample 50 extra points, which led to 150 points, 50 less than in the previous section, reducing the running time and resources. \\
In Figure \ref{AL1}, the distribution of true, picked and posterior are plotted together to estimate how well the two distributions of points match. The posterior distribution results from including the information of the picked points in the GP in the same 200 LHS points of the true distribution. Figure \ref{AL1} shows how these two distributions overlap, meaning the posterior trained in 150 can capture the overall structure of the 200 LHS computed points.
In Figure \ref{AL2}, we can also see how the correction between the posterior and the true values of the 200 LHS points is quite accurate, in concrete 0.887 with an accuracy of 192/200 of well-predicted vulnerable future cases. 
Finally, Algorithm 2 was performed the same manner as PRIM-AL: 100 LHS sampled points were selected at the beginning, and 50 iterations of the sampling algorithm were performed to select 50 extra points.
The comparison with the true distribution of points can be seen in Figure \ref{AL3}. The correlation between the posterior and the true sampled points is 0.894, shown in Figure \ref{AL4} and the accuracy is 188/200 for well-predicted vulnerable/no vulnerable scenarios.
Both algorithms seem to perform similarly well in this proof-of-concept case, which makes us think that sampling points from the borders can capture more or less the same information as sampling them from inside the PRIM box, at least in this proof-of-case study. More studies are needed to generalize these results. On the other hand, the benefit of AL is clear in that it increases data efficiency since fewer data points are needed to archieve similar distributions.
\section{Conclusions}
The use of discovery scenarios in urban planning represents a critical methodological advancement for managing uncertainty and complexity in city development. By integrating exploratory scenario techniques into policy frameworks, planners can enhance urban resilience, foster technological innovation, and develop adaptive governance strategies. This paper describes the first application of Scenario Discovery for the study of the relationship between urban features and individual stress during urban walks.

Key findings from the case study presented in this paper indicate that most of the future scenarios where there is an increase in building density and a rise in the number of individuals with high extraversion (a personality trait) do not lead to lower stress levels, given the proposed vegetation policy in the majority of selected paths in Greater Copenhagen.

Moreover, the amount of vegetation required to reduce stress across all considered future scenarios varies based on the characteristics of each path. Consequently, we showed that a more tailored policy study could be applied both to reduce the resources needed in some areas and to maximise the efficiency of the chosen ones.

Finally, we propose two sampling strategies to enhance the efficiency of the Scenario Discovery process by reducing the number of model runs required, while still capturing the same distribution of vulnerable future scenarios. A proof-of-concept is applied to the Nørrebro path, where we successfully recovered the distribution of vulnerable cases using fewer sampling points, thus decreasing computation time and resource use.

%\subsection{Limitations}

Our research also presents some limitations. Firstly, the small size of the Lisbon dataset made it impossible to achieve a higher accuracy for our stress predictor model, thus causing some deviation in the predictions and potentially masking or changing the impact of some features in the predictions.

Another point is that the data collected in Copenhagen is slightly different from the Lisbon videos, mainly in the people percentage, since the Mapillarity images tend to avoid having people on them. This bias has been addressed by manually increasing the percentage of people by some fixed percentage. Moreover, data from Copenhagen lacks individual characteristics, and the mean value from Lisbon is used for the baseline case in Copenhagen. Ideally, both datasets would have come from the same source type. Also, the coordinate sampling process, a 50mx50m grid, lacks validation of a good representation of each path as a whole.

Finally, we presented a proof-of-concept for the accuracy of the active learning strategies in capturing the distribution of vulnerable points while reducing the needed sample data points. However, our model is relatively simple, and a wider variety of simulators must be tested to prove that the presented algorithms accomplish their goal of successfully recovering the distribution of vulnerable cases with fewer data points under a variety of circumstances. Therefore, further research should be conducted to test different simulations successfully to prove the real efficiency of the proposed algorithms.

In summary, our research underscores the potential of Scenario Discovery as a valuable tool for policymakers, enabling more targeted urban interventions by identifying key stress-related factors. By integrating algorithmic sampling strategies, policy design can become more efficient, reducing computational costs while still capturing critical vulnerabilities. The findings from the case-study emphasize the need for adaptive, locally tailored policies that leverage Scenario Discovery to create stress-reducing urban spaces, moving beyond traditional “what-if” analyses to incorporate uncertainties in decision-making.

\subsection*{Acknowledgements}
We thank eMOTIONAL Cities Grant agreement ID: 945307, funding from EU’s Horizon 2020.4 \footnote{https://emotionalcities-h2020.eu/}
\newpage
\renewcommand\refname{References}

\end{document}